\newcommand{\be}{\begin{equation}}
\newcommand{\ee}{\end{equation}}
\newcommand{\ba}{\begin{eqnarray}}
\newcommand{\ea}{\end{eqnarray}}
\newcommand{\n}{\nonumber \\}

\newcommand{\eq}[1]{(\ref{#1})}
\newcommand{\ad}[1]{{a^\dagger_{#1}}}
\newcommand{\Ad}{A^\dagger}

\newcommand{\z}{t}      

\newcommand{\tpsi}{J}   
\renewcommand{\varphi}{\widehat J}

\documentstyle[12pt]{article}
\begin{document}

\begin{titlepage}
\nopagebreak
\begin{flushright}
November 1994\hfill
RIMS-997\\
YITP/U-94-29\\
SULDP-1994-8\\
UT-693\\
hep-th/9411053
\end{flushright}

\renewcommand{\thefootnote}{\fnsymbol{footnote}}
\vfill
\begin{center}
{\Large Collective Field Theory, Calogero-Sutherland Model}
\vskip 2mm
{\Large and Generalized Matrix Models}

\vskip 20mm

{\large H.~Awata\footnote{JSPS fellow}\setcounter{footnote}
{0}\renewcommand{\thefootnote}{\arabic{footnote}}\footnote{
Research Institute for Mathematical Sciences, Kyoto University,
Kyoto 606, Japan; E--mail:awata@kurims.kyoto-u.ac.jp},
Y.~Matsuo\footnote{Uji research center,
Yukawa Institute for Theoretical Physics,
Kyoto University, Uji 611, Japan; E--mail: yutaka@yukawa.kyoto-u.ac.jp},
S.~Odake\footnote{
Department of Physics, Faculty of Liberal Arts,
Shinshu University, Matsumoto 390, Japan;
E--mail: odake@yukawa.kyoto-u.ac.jp}
and J.~Shiraishi$^*$\footnote{
Department of Physics, University of Tokyo, Tokyo 113, Japan;
E--mail: shiraish@danjuro.phys.s.u-tokyo.ac.jp}
}
\end{center}
\vfill

\begin{abstract}
On the basis of the collective field method, we analyze
the Calogero--Sutherland model (CSM) and the Selberg--Aomoto integral,
which defines, in particular case, the partition function of the
matrix models.
Vertex operator realizations for some of the eigenstates
(the Jack polynomials) of the CSM Hamiltonian are obtained.
We  derive Virasoro constraint for the generalized matrix models
and indicate relations with the CSM operators.
Similar results are presented for the $q$--deformed case
(the Macdonald operator and polynomials), which gives
the generating functional of infinitely many conserved charges in the CSM.
\end{abstract}


\vfill
\end{titlepage}


\section{Introduction}


The purpose of this letter is to discuss some common properties
which are shared by the Calogero--Sutherland model (CSM)
\cite{rCS} described by the Hamiltonian and momentum,
\be
  \widetilde{\cal H}=\frac{1}{2}\sum_{j=1}^N
  \biggl(\frac{1}{i}\frac{\partial}{\partial q_j}\biggr)^2
  +\frac{1}{2}\Bigl(\frac{\pi}{L}\Bigr)^2
  \sum_{i\neq j}\frac{\beta(\beta-1)}{\sin^2\frac{\pi}{L}(q_i-q_j)},
  \quad
  \widetilde{\cal P}=\sum_{j=1}^N
  \frac{1}{i}\frac{\partial}{\partial q_j},
  \label{e1}
\ee
and the ``generalized matrix model'' whose partition function
is defined by the following integral,
\be
  Z_{\beta}([g])\equiv \int \prod_{i=1}^{M} dt_i
  \left|\Delta(t)\right|^{2\beta}
  e^{\sum_{n=0}^\infty g_{n} \sum_{i=1}^{M} t_i^n},
  \quad
  \Delta(t)\equiv\prod_{i<j} (t_i-t_j).
  \label{e2}
\ee
For some specific values of $\beta$, the latter integral is
related to the usual matrix models\cite{rMat1};
the hermitian matrix model ($\beta=1$)\cite{rMat2},
the orthogonal matrix model
($\beta=\frac{1}{2}$)\cite{rMat3}
and the symplectic matrix model ($\beta=2$)\cite{rMat4}.
The integral of this type, called the Selberg--Aomoto integral,
was studied in \cite{rK}
as a multivariate generalization of the hypergeometric integral
and has been used recently for calculating
the correlation functions of the CSM \cite{rF}\cite{rLPS}.

In the present letter, we apply the collective field methods
\cite{rJS}\cite{rAJL}\cite{rCF1}\cite{rCF2} to these two models.
Firstly, in the CSM, by using a collective field Hamiltonian,
we derive some of the eigenstates as the vertex operators.
Mathematically, they are known as Jack symmetric polynomials \cite{rS}
and are classified by the Young diagrams.
Furthermore, 
the mutually commutative conserved charges of the CSM are known to be
realized as the Cartan generators of the $W_{1+\infty}$ algebra \cite{rHW}.
These properties strongly indicates that the system has the $W_{1+\infty}$
symmetry \cite{rWinf}\cite{rKR}\cite{rAFMO}.

Secondly, we study the integral \eq{e2}
and show the appearance of the Virasoro constraint \cite{rSD}.
Although they are defined by the Virasoro generators with the mode $n\geq -1$,
they have a unique relativistic extension
with the following central charge,
\be
c=1-\frac{6(1-\beta)^2}{\beta}.
\label{e3}
\ee
This formula satisfies the duality symmetry,
$\beta\leftrightarrow \frac{1}{\beta}$,
which is the known property of the Jack polynomials \cite{rS}.

The CSM and the Selberg--Aomoto integral might be related by
the integral representation of the Jack polynomials.
We show, through this relation, the vertex operators used in the CSM
are the primary fields with dimension 1, {\it i.e.} the screening currents.

Finally, we refer to a $q$--deformation of the foregoing discussions.
Mathematically, the corresponding
Hamiltonian operator and the eigenstates are known
as the Macdonald operator and the Macdonald polynomials.
They are useful in obtaining the higher conservative charges of the CSM.


\section{Hamiltonian and Collective coordinates}


To fix some notations, we start from summarizing 
the collective field description of the CSM.
Let us make the coordinate transformation,
$x_j\equiv e^{2\pi iq_j/L}$.
The eigenfunctions $\psi_\lambda(x)$ of the CSM are then factorized as,
\ba
  \psi_\lambda(x)
  &\!\!=\!\!&
  \tpsi_\lambda(x) \widetilde{\Delta}(x)^\beta, \n
  \widetilde{\Delta}(x)
  &\!\!=\!\!&
  \prod_{i<j}\sin\frac{\pi}{L}(q_i-q_j)
  \propto\prod_i x_i^{-(N-1)/2}\prod_{i<j}(x_i-x_j),
  \label{e4}
\ea
where
$\tpsi_\lambda(x)$
is the symmetric polynomial of the coordinates $x_i$
$(i=1,2,\cdots,N)$.
The Hamiltonian itself is modified when it is acted on
$\tpsi_\lambda(x)$,
\ba
  &&
  \widetilde{\Delta}(x)^{-\beta}\widetilde{\cal H}
  \widetilde{\Delta}(x)^{\beta}
  =2\Bigl(\frac{\pi}{L}\Bigr)^2 {\cal H}+E_0, \n
  &&
  {\cal H}\equiv
  \sum_{i=1}^N D_i^2
  +\beta\sum_{i<j}\frac{x_i+x_j}{x_i-x_j}(D_i-D_j),
\label{e5}
\ea
where $D_i\equiv x_i\frac{\partial}{\partial x_i}$ and
$E_0=\frac{1}{6}(\frac{\pi}{L})^2\beta^2(N^3-N)$
is the eigenvalue of the groundstate $\widetilde{\Delta}(x)^{\beta}$.
Similarly, the momentum is modified as,
\be
  \widetilde{\Delta}(x)^{-\beta}\widetilde{\cal P}
  \widetilde{\Delta}(x)^{\beta}
  = 2\frac{\pi}{L}{\cal P},\qquad
  {\cal P}\equiv\sum_{i=1}^ND_i.
\label{e6}
\ee

Eigenfunctions $\tpsi_\lambda(x)$ of ${\cal H}$
are known as the Jack symmetric
polynomials \cite{rS}.
They are parametrized by the Young diagrams and the
eigenvalue associated with the diagram
$\lambda=(\lambda_1,\cdots,\lambda_M)$
is given by,
\be
\epsilon_{\lambda}=
  \sum_{i=1}^M\Bigl(\lambda_i^2+\beta(N+1-2i)\lambda_i\Bigr)
= \sum_{i=1}^{M'}\Bigl(-\beta\lambda_i^{\prime 2}
                       +(\beta N+2i-1)\lambda_i'\Bigr),
\ee
where ${}^t\lambda=(\lambda_1',\cdots,\lambda_{M'}')$
is the conjugate of $\lambda$.


Because $\tpsi_\lambda(x)$ is symmetric
with respect to $x_i$'s,
it can be written out by using the power sum polynomials
$p_n \equiv \sum_ix_i^n$.
Let us denote $\ad{n}$ as the ``creation operator''
which gives rise to this $p_n$.  The ``annihilation operator''
associated with it is defined by the commutation relation,
\be
  \Bigl[a_n,\ad{m}\Bigr]=\frac{1}{\beta}n\delta_{n,m},
  \quad (n,m>0).
\label{e7}
\ee
We introduce the vacuum states by
$a_n |0\rangle=\langle 0|\ad{n}=0$ ($n>0$).
One may translate the collective field Hilbert
space and the space of symmetric polynomials
by the formula,
\be
\langle 0|e^{\beta \sum_i A(x_i)}
\ad{n_1}\cdots \ad{n_m}|0\rangle
=  p_{n_1}\cdots p_{n_m},\qquad
A(x)\equiv \sum_{n=1}^\infty \frac{1}{n} a_n x^n.
\label{e11}
\ee

These operators are related with the conventional collective
field operators,
$\rho(x)= \sum_{i=1}^N \delta(x-x_i)$,
by
$\int dx \,x^n\rho(x) = p_n$.
Use of
these combinations has the benefit to
illuminate the relation with the matrix--type integral
\eq{e2} more directly.

The convention of the definition of \eq{e7}
is to reproduce the standard inner product between the
symmetric polynomials \cite{rS}\cite{rLPS}, {\it i.e.},
\ba
  \langle p_1^{r_1}\cdots p_n^{r_n},
  p_m^{s_m}\cdots p_1^{s_1}\rangle
  &\!\!\equiv \!\!&
  \langle 0|a_1^{r_1}\cdots a_n^{r_n}
  \ad{m}^{\!\!s_m}\cdots\ad{1}^{s_1}|0\rangle, \n
  &\!\!=\!\!&
  \delta_{\{r\},\{s\}}\beta^{-\sum_{i=1}^nri}
  \prod_{i\geq 1}i^{r_i}r_i!.
\ea

One may derive the collective coordinate representation
of the Hamiltonian and the momentum in a usual way.
The Hamiltonian in terms of the collective coordinates,
$\widehat{\cal H}$,
is defined by the transformation in \eq{e11} as,
\be
{\cal H}\langle 0|e^{\beta \sum_i A(x_i)}
=\langle 0|e^{\beta \sum_i A(x_i)}\widehat{\cal H}.
\label{e12}
\ee
For example, the momentum operator is obtained as follows,
$$
\sum_{i}D_i\langle 0|e^{\beta \sum_i A(x_i)}
= \sum_i \langle 0|e^{\beta \sum_i A(x_i)}\beta\sum_{n=1}^\infty x_i^n a_n=
 \langle 0|e^{\beta \sum_i A(x_i)}\sum_n(\beta \ad{n} a_n),
$$
which gives, $\widehat{\cal P}=\beta\sum_{n=1}^\infty \ad{n}a_n.$
Similarly the Hamiltonian is given by,
\ba
  \widehat{\cal H}
  &\!\!=\!\!&
  \beta^2 \sum_{n,m=1}^{\infty}
  (\ad{n+m}a_na_m+\ad{n}\ad{m}a_{n+m}) \n
  &&
  +\beta(1-\beta)\sum_{n=1}^{\infty}n\ad{n}a_n
  +\beta^2N\sum_{n=1}^{\infty}\ad{n}a_n.
\ea
This particular form of the Hamiltonian appeared previously, for
example,  in \cite{rCF1} (see also the recent work \cite{rISO}).


\section{Vertex operators and Energy eigenstates}


Some of the simpler eigenfunctions (the Jack polynomials) of the
Hamiltonian can be explicitly written down by using the vertex operators.
As is known, they are parametrized by the Young diagrams.
In this section, we give their bosonized forms
when the Young diagram has,
(i) one or two rows (columns) or
(ii) one row and one column (one ``hook'').

Define,
$\Ad(\z)\equiv \sum_{n=1}^\infty \frac{1}{n} \ad{n} \z^n.$
There are two types of the vertex operators
which are diagonalized by the action of the Hamiltonian
$\widehat{\cal H}$ as,
\be
  \widehat{\cal H} e^{\gamma\Ad(\z)}|0\rangle
  =
  \biggl(\frac{\beta}{\gamma}D^2 +(\beta N-\gamma)D\biggr)
  e^{\gamma\Ad(\z)}|0\rangle,
  \label{e14}
\ee
where $\gamma=\beta,-1$ and $D=\z\frac{\partial}{\partial \z}$.
Therefore, by expanding these vertex operators in terms of $\z$,
\be
  e^{\gamma \Ad(\z)} 
  =
  \sum_{n=0}^{\infty}\varphi_n^{(\gamma)}(\ad{}) \z^n ,
\label{e15}
\ee
$\varphi_n^{(\beta)}(\ad{})|0\rangle$ and
$\varphi_n^{(-1)}(\ad{})|0\rangle$ become
the eigenstates of the Hamiltonian.  Indeed, the symmetric
polynomials associated with them are the Jack polynomials
of the Young diagram with single row $(n)$ and single column $(1^n)$,
respectively.

For the state $\prod_{i=1}^Me^{\gamma\Ad(\z_i)}|0\rangle$, we have,
\ba
  \widehat{\cal H}\prod_{i=1}^Me^{\gamma\Ad(\z_i)}|0\rangle
  &\!\!=\!\!&
  \biggl(\sum_{i=1}^M\Bigl(\frac{\beta}{\gamma}D_i^2
  +(\beta N-\gamma)D_i\Bigr) \n
  &&
  \;\;+2\gamma\sum_{i<j}\frac{1}{1-\frac{\z_j}{\z_i}}
  \Bigl(\frac{\z_j}{\z_i}D_i-D_j\Bigr)\biggr)
  \prod_{i=1}^Me^{\gamma\Ad(\z_i)}|0\rangle,
  \label{HVV}
\ea
where $D_i=\z_i\frac{\partial}{\partial \z_i}$
and $|\z_j/\z_i|<1$ for $i<j$.
In the case of $\beta=1$ when the Jack polynomial reduces to the
Schur polynomial, the eigenstates of $\widehat{\cal H}$ are given by
$\oint\prod_{j=1}^M\frac{d\z_j}{2\pi i}\z_j^{-\lambda_j-1}
\prod_{i<j}(1-\frac{\z_j}{\z_i})\prod_{i=1}^Me^{\pm\Ad(\z_i)}|0\rangle$,
which can be rewritten in a determinant form,
$\det (\varphi_{\lambda_i-i+j}^{(\gamma)})_{1\leq i,j\leq M}|0\rangle$.
For $\beta\neq 1$, however, this is no longer the case.
Expanding \eq{HVV}, we obtain,
\ba
  &&
  \widehat{\cal H}
  \varphi_{n_1}^{(\gamma)}\cdots\varphi_{n_M}^{(\gamma)}|0\rangle \n
  &\!\!=\!\!&
  \sum_{i=1}^M\biggl(\frac{\beta}{\gamma}n_i^2
  +\Bigl(\beta N-\gamma(2i-1)\Bigr)n_i\biggr)
  \varphi_{n_1}^{(\gamma)}\cdots\varphi_{n_M}^{(\gamma)}|0\rangle \n
  &&
  +2\gamma\sum_{i<j}\sum_{r=1}^{n_j}(n_i-n_j+2r)
  \varphi_{n_1}^{(\gamma)}\cdots\varphi_{n_i+r}^{(\gamma)}\cdots
  \varphi_{n_j-r}^{(\gamma)}\cdots\varphi_{n_M}^{(\gamma)}|0\rangle.
\ea
At this moment, we are successful in diagonalizing
this equation only for the cases $M=1,2$.
For $M=2$ case,
the eigenstates  correspond to the Young diagram with two rows
($\gamma=\beta$) or two columns ($\gamma=-1$).
The explicit form of the diagonalized basis are given by,
\ba
  \varphi_{(\lambda_1,\lambda_2)}^{(\gamma)}(\ad{})|0\rangle
  &\!\!=\!\!&
  \sum_{\ell=0}^{\lambda_2}c^{(\gamma)}(\lambda_1-\lambda_2,\ell)
  \varphi_{\lambda_1+\ell}^{(\gamma)}(\ad{})
  \varphi_{\lambda_2-\ell}^{(\gamma)}(\ad{})
  |0\rangle, \n
  c^{(\gamma)}(\lambda,\ell)
  &\!\!=\!\!&
  \frac{\lambda+2\ell}{\lambda+\ell}
  \prod_{j=1}^{\ell}\frac{\lambda+j}{j}\cdot
  \prod_{i=1}^{\ell}\frac{-\gamma+\frac{\beta}{\gamma}(i-1)}
                         { \gamma+\frac{\beta}{\gamma}(\lambda+i)}.
\ea



One can easily show that
the Jack polynomials of single hook $(n,1^m)$ are,
\be
  \varphi_{(n,1^m)}(\ad{})|0\rangle =
  \sum_{\ell=0}^{m} \left(n+\ell+\beta(m-\ell\:)\right)\:
  \varphi_{n+\ell}^{(\beta)}(\ad{})
  \varphi_{m-\ell}^{(-1)}(\ad{})  |0\rangle.
\ee


\section{Virasoro constraint}


Let us go back to the Selberg--Aomoto integral \eq{e2}.
The collective field method can be also applied
here.  Namely, the insertion of the operator,
$\sum_{i=1}^{M} t_i^n$, can be realized by
taking a partial derivative with respect to the coupling $g_n$.
These operators can be combined
to give a single free collective field,
\be
\partial \phi(z) = \sqrt{2\beta} \sum_{n=0}^\infty
\frac{\partial}{\partial g_n} z^{-n-1}+ \frac{1}{\sqrt{2\beta}}
\sum_{n=1}^\infty ng_n z^{n-1}.
\label{e19}
\ee
The coefficients for the bosonic field
is chosen for the later convenience.

The essential feature of the matrix--type integral can
be extracted by considering a set of differential equations.
For the hermitian matrix case,
it is so--called the Virasoro constraint \cite{rSD}.
We may derive similar equations for our generalized integral \eq{e2}.
The method we employ here is essentially the same as the
hermitian case, namely we start from the integral which
is trivially zero and rewrite it as the differential operator
of the source term acting on the original integral,
\ba
  0
  &\!\!=\!\!&
  \int \prod_{i=1}^{M}dt_i
  \sum_{i=1}^{M} \frac{\partial}{\partial t_i}
  \biggl(t_i^{n+1}\left|\Delta(t)\right|^{2\beta}
  e^{\sum_{\ell=0}^\infty g_\ell \sum_{i=1}^{M} t_i^\ell}\biggr) \n
  &\!\!=\!\!&
  L_n Z([g]),  \qquad (n=-1,0,1,2,\ldots), \\
  L_n
  &\!\!=\!\!&
  \sum_{m=1}^\infty mg_m\frac{\partial}{\partial g_{n+m}}
  +\beta\sum_{m=0}^n
  \frac{\partial^2}{\partial g_m\partial g_{n-m}}
  +(1-\beta)(n+1)\frac{\partial}{\partial g_n}.
\label{e20}
\ea

The Virasoro generators appearing here have the mode $n$ greater than $-1$.
Hence there is no central extension in the commutation
relations between these operators.
However, we may uniquely extend them as the components
of the relativistic energy--momentum tensor,
\be
  T(z)= \sum_{n=-\infty}^\infty L_n z^{-n-2}
  = \frac{1}{2}:(\partial \phi(z))^2:
  -\frac{1-\beta}{\sqrt{2\beta}}\partial^2\phi(z).
\label{e21}
\ee
This energy--momentum tensor satisfies
the Virasoro algebra with central charge \eq{e3}.

At this moment, the physical meaning of this
central charge is obscure.  For the hermitian case,
the double scaling limit is described by the KP--hierarchy.
The partition function is identified as the
$\tau$--function.  Since KP--hierarchy is essentially
the free fermion system with $c=1$, the central
charge \eq{e3} looks plausible.
The nontrivial values for other matrix models,
(for orthogonal or symplectic case, $c=-2$),
may indicate that the double scaling limit
for those models is described by
interacting system.


\section{Integral representation and Virasoro symmetry}


We now present the relations between
two models defined by \eq{e1} and \eq{e2}.
The original Selberg--Aomoto integral studied in \cite{rK} is as follows,
\ba
\widetilde S_{M,N}(\lambda_1,\lambda_2,\lambda, \mu;
[x])&\!\!=\!\!&
\int_{\left[ 0,1 \right]^{M}} \prod_{i=1}^{M} dt_i \cdot
\prod_{i=1}^{M} \prod_{k=1}^{N}
(1-t_i x_k)^\mu D_{\lambda_1,\lambda_2,\lambda}([t]),\n
 D_{\lambda_1,\lambda_2,\lambda}([t])&\!\!=\!\!&
\prod_{i=1}^{M} t_i^{\lambda_1} (1-t_i)^{\lambda_2}
\prod_{i < j}|t_i-t_j|^\lambda.
\label{eSAI}
\ea
This integral satisfies the multivariate generalization 
of the hypergeometric differential equation when $\mu=-\lambda/2$ or $1$.
Furthermore, it can be expanded by Jack polynomials with
$\beta=2\mu^2/\lambda$
in the similar way as the Taylor expansion of the hypergeometric function.
This fact has been used \cite{rF} for discussing
the correlation functions of the CSM.

The correspondence between \eq{eSAI} and our integral \eq{e2}
is given when we make the transformation of variables,
{\it i.e.} from $x_k$ to $p_n=\sum_{k=1}^{N} x_k^n$,
\be
\prod_{i=1}^{M} \prod_{k=1}^{N} (1-t_i x_k)^\mu=
\prod_{i=1}^{M} e^{-\mu\sum_{n=1}^\infty\frac{1}{n}  p_n t_i^n}.
\ee
The ``Vertex operators'' which appear on the right hand
side are exactly the same as those which appeared in \eq{e14}.

The Virasoro symmetry considered in the previous section
is related to the CSM in this context.
Indeed,
\be
\widehat{\cal H}=\beta\sum_{n=1}^\infty \ad{n} L_n
+ ( \beta(N+1-2a_0)-1) \widehat{\cal P},
\label{e22}
\ee
where $a_n=\frac{\partial}{\partial g_n}$ for $n\geq 0$
and $\ad{n}= \frac{n}{\beta}g_n$ for $n>0$.
Furthermore, the vertex operators \eq{e14} are
nothing but the screening currents in terms of the Virasoro
generators,
\be
  L_n e^{\gamma\Ad(\z)}|0\rangle
  =
  \partial_\z\Bigl(\z^{n+1}e^{\gamma\Ad(\z)}\Bigr)|0\rangle.
\ee
for $n\geq -1$.

Although linear combinations of the Jack polynomials
can be obtained by the integral \eq{eSAI},
it is interesting if one may derive the direct
integral representation of the Jack polynomials.
However, it is still difficult to
find the general form of such integral representation,
some of the simpler ones can be written as follows,
\be
  \oint\prod_{j=1}^M\frac{dt_j}{2\pi i}
  \prod_{i=1}^M\prod_{k=1}^N(1-t_i x_k)^{-\gamma}
  \prod_{i<j}(t_i-t_j)^{2\frac{\gamma^2}{\beta}}
  \prod_{i=1}^M t_i^{-\lambda_i-1-\frac{\gamma^2}{\beta}(M-1)},
\label{eRA}
\ee
where,
\be
  \lambda_i=
  \left\{\begin{array}{ll}
         \lambda+1&(1\leq i\leq m)\\
         \lambda  &(m+1\leq i\leq M)
         \end{array}
  \right.,\quad
  0\leq m\leq M-1.
\ee
These Jack polynomials correspond to the Young diagram
$\lambda=(\lambda_1,\cdots,\lambda_M)$ and
${}^t\lambda=(\lambda_1,\cdots,\lambda_M)$
for $\gamma=\beta$ and $-1$, respectively.


\section{$q$-Deformation and Macdonald polynomials}


Finally, we briefly discuss 
the $q$--deformation and the Macdonald polynomials $Q_{\lambda}$ \cite{rMac}
by using the method developed in the previous sections.
The detail will appear elsewhere.
The Macdonald operator, 
\be
  D_{q,t}=\sum_{i=1}^N \prod_{j\neq i}\frac{t x_i-x_j}{x_i-x_j} T_{q,x_i},
\ee
plays the same role as the CSM Hamiltonian ${\cal{H}}$,
where $T_{q,x_i}$ is the $q$-shift operator,
\be
T_{q,x_i}f(x_1, \cdots, x_N) = f(x_1,\cdots,qx_i,\cdots,x_N).
\ee
Here, a new complex deformation parameter $q$ is introduced and
$t$ is related to $\beta$ by $t=q^{\beta}$.

The situation in previous sections can be obtained by taking the limit
$\hbar \rightarrow 0$ with $q=e^\hbar$.
In this limit, the Macdonald operator behaves as
$D_{q,t}=\sum_{n\ge 0} D_{q,t}^{(n)} {\hbar^n}/{n!}$
with,
\ba
D_{q,t}^{(0)}&\!\!=\!\!&N,\qquad
D_{q,t}^{(1)}= {\cal P} +\frac{\beta}{2}N(N-1),\n
D_{q,t}^{(2)}&\!\!=\!\!& {\cal H}+\beta(N-1) {\cal P}
+\frac{\beta^2}{6}N(N-1)(2N-1).
\label{eEXPANSION}
\ea
here ${\cal H}$ and ${\cal P}$ are in \eq{e5} and \eq{e6}.
In this sense, the Macdonald operator can be regarded as the
generating functional of the infinitely many conserved charges
in the CSM.

Amazingly, one may find a closed form of
collective field representation for
the Macdonald operator $D_{q,t}$ as follows,
\ba
  &\!\!\!\!& D_{q,t}\langle 0|
  e^{\sum_{n=1}^\infty \frac{1-t^n}{1-q^n}\frac{a_n}{n} p_n}
  =  \langle 0|
  e^{\sum_{n=1}^\infty \frac{1-t^n}{1-q^n}\frac{a_n}{n} p_n}
   \widehat{D}_{q,t},\n
  &\!\!\!\!& \widehat{D}_{q,t}= \frac{t^{N}}{t-1}
    \oint \frac{dz}{2\pi i} \frac{1}{z}
e^{ \sum_{n=1}^\infty \frac{1-t^{-n}}{n} \ad{n} z^n   }
e^{-\sum_{n=1}^\infty \frac{1-t^{ n}}{n}  a_{n} z^{-n}}
-\frac{1}{t-1},
\label{eMcCF}
\ea
where the commutation relations for the bosonic oscillators are deformed as
$[a_n,a^{\dagger}_m]= n \frac{1-q^{n}}{1-t^{n}}\delta_{n,m}$.
Expanding this expression, we find the bosonized momentum and
Hamiltonian as the coefficients of $\hbar$ and $\hbar^2$.

Similar to our discussion in section 3,
we obtain the bosonized realization for some of the
eigenstates (the Macdonald polynomials) of the Macdonald operator.
Let,
\be
  \exp\left( \sum_{n=1}^\infty \frac{1-q^{\gamma n}}{1-q^n}
          \frac{a_{n}^{\dagger} }{n} z^n\right) 
  =
  \sum_{n=0}^{\infty} \widehat Q_n^{(\gamma)} z^n 
\ee
the states $\widehat Q_n^{(\gamma)}|0\rangle$ with $\gamma=\beta$ or $-1$
are the Macdonald polynomials corresponding to the Young diagram
with single row $(n)$ or single column $(1^n)$, respectively.
That of 
two rows $\lambda=(\lambda_1,\lambda_2)$ or
two columns ${}^t\lambda=(\lambda_1,\lambda_2)$ are given by,\footnote{
A conjecture for the special case of this expression was
derived during the discussion with
H. Kubo. We understand that this result is independently
obtained by A. N. Kirillov.}
\ba
  \widehat Q_{(\lambda_1,\lambda_2)}^{({\gamma})}|0\rangle
  &\!\!=\!\!&
  \sum_{\ell=0}^{\lambda_2}c^{({\gamma})}(\lambda_1-\lambda_2,\ell)
  \widehat Q_{\lambda_1+\ell}^{({\gamma})}
  \widehat Q_{\lambda_2-\ell}^{({\gamma})}
  |0\rangle, \n
  c^{({\gamma})}(\lambda,\ell)
  &\!\!=\!\!&
  \frac{1-q^{\frac{\beta}{\gamma}(\lambda+2\ell)}}
       {1-q^{\frac{\beta}{\gamma}(\lambda+\ell)}}
  \prod_{j=1}^{\ell}
   \frac{1-q^{\frac{\beta}{\gamma}(\lambda+j)}}
        {1-q^{\frac{\beta}{\gamma}j}}\cdot
  \prod_{i=1}^{\ell}
  \frac{q^{\gamma}-q^{\frac{\beta}{\gamma}(i-1)}}
       {1-q^{\gamma+\frac{\beta}{\gamma}(\lambda+i)}},
\ea
with $\gamma=\beta$ or $-1$, respectively.
 The Macdonald polynomials of single hook $(n,1^m)$ are,
\be
  \widehat Q_{(n,1^{m})}|0\rangle =
  \sum_{\ell=0}^{m} \frac{1-q^{n+\ell}t^{m-\ell}}{1-q} q^{m-\ell}\:
  \widehat Q_{n+\ell}^{(\beta)}\widehat Q_{m-\ell}^{(-1)}|0\rangle.
\ee
%
%

\vskip 5mm

\noindent{\bf Acknowledgments:}


We would like to thank M.~Fukuma, T.~Inami, S.~Iso, N.~Kawakami,
A.~N.~Kirillov, H.~Kubo, M.~Ninomiya and P.~Wiegmann for discussions and
encouragements. Some parts of this work were done in Hotaka-Sanso.
This work is supported in part by Grant--in--Aid for Scientific
Research from Ministry of Science and Culture.


\noindent{\bf Note Added:}

After we submitted this paper,
we learned that Avan and Jevicki \cite{rAJ} discussed the connection
between $c=1$ matrix model and the Calogero Moser model.
In this context, we have to  mention the work by
Simons et. al. \cite{rSLA} where the matrix model technique
was used to derive  two-point correlation functions for
the CSM with
$\beta=1,2,\frac{1}{2}$.
We were also indicated that
G. Harris \cite{rH} studied the Virasoro
constraint of the matrix
model for non-orientable surfaces.
Although the purpose of these works
is different from ours, i.e. to relate the CSM
with the conformal field theory
with $c<1$, they  give  complementary viewpoints to the problem.

As for the approach which is the closest to ours,
we would like to mention the recent announcement by
Mimachi and Yamada \cite{rMY} where they expressed
the Virasoro singular vectors as
the Jack polynomials of rectangular Young diagrams,
i.e. our equation \eq{eRA}.


\end{document}